\documentstyle[12pt,aasms]{article}

\def\fun#1#2{\lower0.837ex\vbox{\baselineskip0ex\lineskip0.209ex
  \ialign{$\mathsurround=0ex#1\hfil##\hfil$\crcr#2\crcr\sim\crcr}}}

\def\msun{M_\odot}
\def\msunyr{M_\odot \ {\rm yr}^{-1}}
\def\sles{\lower2pt\hbox{$\buildrel {\scriptstyle <}
   \over {\scriptstyle\sim}$}}
 
\def\sgreat{\lower2pt\hbox{$\buildrel {\scriptstyle >}
   \over {\scriptstyle\sim}$}}

\begin{document}

 \title{ The October 1985 Long Outburst of U Geminorum:
   Revealing the Viscous Time Scale
 in Long Orbital Period Dwarf Novae}

 \vskip 1truein

 \author{John K. Cannizzo\footnote{
also University of Maryland    Baltimore County}}
 \affil{e-mail: cannizzo@stars.gsfc.nasa.gov}
 \affil{NASA/GSFC/Laboratory for High Energy Astrophysics, 
 Code 661, Greenbelt, MD 20771}
 \authoraddr{NASA/GSFC/Laboratory for High Energy Astrophysics, 
 Code 661, Greenbelt, MD 20771}

\medskip
\medskip
\medskip

 \author{ Neil Gehrels }
 \affil{e-mail: gehrels@lheapop.gsfc.nasa.gov}
 \affil{NASA/GSFC/Laboratory for High Energy Astrophysics, 
 Code 661, Greenbelt, MD 20771}
 \authoraddr{NASA/GSFC/Laboratory for High Energy Astrophysics, 
 Code 661, Greenbelt, MD 20771}

\medskip
\medskip
\medskip

 \author{ Janet A. Mattei} 
 \affil{e-mail:  jmattei@aavso.org }
 \affil{
 American Association of Variable Star Observers,
 25 Birch Street,
 Cambridge, MA 02138-1205 }
 \authoraddr{
 American Association of Variable Star Observers,
 25 Birch Street,
 Cambridge, MA 02138-1205 }

\medskip
\medskip
\medskip

%\vskip 3truein
%\vskip 0.25truein

%\vskip 0.5truein
%\centerline{submitted to the Astrophysical Journal (Letters)} 
%  1998, February 10, vol. 494}

%\received{ 2001 September 7}
%\accepted{ 2001 October 1}

\begin{abstract}

We examine the AAVSO light curve of U Geminorum
 from 1908 to 2002, with particular
focus on
the October 1985 outburst.
This outburst was longer than any other seen
in U Gem by about a factor of 2,
and
 appears to be unique among all
dwarf nova outbursts seen in systems
with orbital periods longer than 3 hr
in that one can measure the decay time
scale during the initial slow decay.
This rate 
 is $\sim26\pm6$  d mag$^{-1}$.
  Using estimates of
the rate of accretion during outburst
taken from Froning et al., 
one can show that 
$\sim10^{24}$ g
of gas was accreted  onto the white dwarf
  during the outburst
which
   constrains the surface density
in the outer accretion disk
  to be $\sim600$ g cm$^{-2}$.
 The observed time scale
for decay is consistent 
with that expected in U Gem, 
given its orbital period and disk mass
at the time the outburst began.
 The data are not of
sufficient quality to be able to
 ascertain a 
     deviation from  exponentiality
in the decay light curve (as in the
SU Ursa Majoris stars' superoutbursts).
% as
%is expected qualitatively during a 
%``viscous decay''
%    in which the dominant mode of
%    depletion 
% of the gas stored
%   in the accretion disk
%is accretion onto the central object.
 When coupled with the viscous time
inferred from the (short orbital period) SU UMa stars,
the U Gem viscous time scale lends support
to the standard model in which the decays
in dwarf novae can either be viscous or
thermal, with the  ratio between them
being roughly $h/r$ where $h$ is the vertical
pressure scale height in the disk.
 Indeed, dwarf novae are the only systems
for which one can be reasonably certain of
the identification of ``viscous'' and ``thermal''
decays.

\end{abstract}

\medskip
\medskip

{\it  Subject headings:}
accretion,  accretion disks $-$ binaries: close  $-$
                                   stars: individual (U Gem)

\section{ Introduction }

Cataclysmic variables (CVs)
 are
interacting binaries in which a Roche lobe filling
K or M dwarf secondary transfers matter
onto a white dwarf (WD) primary (Warner 1995). 
Dwarf novae (DNe) make up
a subclass of CVs which is characterized
by semiperiodic outbursts
caused by episodic accretion of stored
gas onto the WD (Meyer \& Meyer-Hofmeister 1981,
see Cannizzo 1993a for a review).
 U Geminorum is one of the two prototypical 
dwarf novae, the other one being SS Cygni.
  U Gem was the first CV to be discovered
(in 1855 by John Russell Hind of Bishop's
Observatory, London), and is among the better studied.
     It has an orbital period of 4.25 hr,
 a primary mass $\sim1.1\msun$, secondary 
mass  $M_{\rm WD}\sim0.4\msun$, and inclination $\sim67^{\deg}$
(Krzeminski 1965, Smak 1971, 1976,
Warner \& Nather 1971,
  Sion et al. 1998, Long \& Gilliland 1999).
Krzeminski (1965) showed U Gem to be an eclipsing binary
system, and Warner \& Nather (1971) and independently Smak (1971)
  developed the now standard model for nonmagnetic CVs,
with U Gem as the prototype.
(The inference on orbital inclination comes from the fact that
 the disk is eclipsed, but not the WD.)
U Gem  also has a reliable distance
  $96.4\pm4.6$ pc
determined from {\it HST}
  astrometric parallax (Harrison et al. 1999).
%   In terms of its outburst characteristics,
The  mean recurrence time for outbursts
 in U Gem 
    is $\sim120$ d,
during which time it brightens
from $m_V\simeq14$ to  $m_V\simeq 9$ (Szkody \& Mattei 1984).

The outburst durations in U Gem follow a bimodal
distribution as with  other DNe.
  The alternation of long and short outbursts
has been shown to be a natural consequence
of the accretion disk limit cycle model
(Buat-Menard et al. 2001).
The   outbursts last typically
either $\sim 3-8$ d or  $\sim12-16$ d (measured
through $m_V =12$ on the rising and decaying branches).
There was, however, one notable exception to this:
in October 1985 a long outburst occurred
which lasted $\sim40$ d.  This long duration
is interesting from the standpoint of 
the accretion disk limit cycle theory which 
provides the theoretical underpinnings for our
current understanding of dwarf nova outbursts.
  According to theory,  a decrease in the light
output from the disk can occur in one of two ways:
(1) One can have the entire disk completely in a
hot, ionized state, and have the mass and concomitant
 mass accretion rate decrease by virtue of loss of
disk material onto the WD $-$ a ``viscous'' decay, or
(2) A cooling front can sweep through the disk from
   outside to inside and revert the disk back to a
   neutral gas, with very low viscosity, that does
   not accrete appreciably $-$ a ``thermal'' decay.
  The ratio of the times  between these two
time scale is roughly $h/r \simeq 0.03$, the ratio
of a local pressure scale height to radius.
 For some reason (which would be difficult to uniquely
identify),
    the cooling front in the Oct 1985 outburst 
  was delayed for much longer than usual, thereby forcing
the entire disk to remain in the hot, ionized state
and to continue accreting slowly onto the WD.

  Although the ``fast'' decays during which the
cooling front traverses the disk have been extensively
studied in dwarf novae, and in fact form the basis
for the ``Bailey''  relation which provides our only direct
measure of the Shakura \& Sunyaev (1973) $\alpha$
parameter for ionized gas,
 the long outbursts seen in the well-studied
long orbital systems such as U Gem and SS Cyg
do not exhibit enough dynamic range in flux
               to be able to determine
 a time scale for the slowly decaying  
     portion of the light curve.
The term ``viscous plateau''
has been used for the slow decays
to convey the fact that the flux
is roughly constant (Cannizzo 1993b).
  For instance, an analysis of the long term AAVSO
light curve of SS Cyg reveals no ``outlier'' outburst
of anomalously long duration (see Fig. 2 of Cannizzo \& Mattei
   1992) like the  Oct 1985 U Gem outburst.
 This is significant  given that the 
Cannizzo \& Mattei study
      covers $\sim10$ times more outbursts than
in the current  U Gem study.
     Only in the ``superoutbursts''
    of the SU UMa subclass of DNe
can the slow decay time scale be reliably determined 
     (e.g., Cannizzo 2001b presents a study of the 2001 
superoutburst of WZ Sge).
    The October 1985 outburst of U Gem
 appears to be the only non-SU UMa-type event ever observed
for which one can also make this determination
and therefore ``test'' (in one limited way)
               the prevailing paradigm for
viscous and thermal decays in accretion disks.

\section { Background }

Froning et al. (2001)
 present {\it FUSE} observations
of U Gem at  three times near
outburst maximum and one near quiescence.
From their spectral fitting they derive a
rate of accretion in the disk
near maximum of $\sim5\times 10^{-9}\msunyr$.
At this rate, a total mass of $\sim 1\times 10^{24}$ g
would be accreted in 35 d, the approximate duration of
the viscous plateau for the outburst of current interest.
If roughly half of the mass in the disk
at the start of the outburst
 is accreted onto the WD during the outburst,
this implies an initial disk mass
 of $\sim2\times 10^{24}$ g.
For an outer disk radius
    (determined by the orbital
period and WD mass) of $4.5\times 10^{10}$ cm,
this gives a surface density at the
outer disk edge of $\sim600$ g cm$^{-2}$
at the start of an outburst,
for a quasi-steady disk in which $\Sigma\propto r^{-3/4}$.

The controlling (i.e., slowest) time scale
in a quasi-steady disk without transition fronts
is the viscous time scale at the outer
edge $\tau_{\nu}(r_{\rm outer})$.
One can obtain a simple estimate of
$\tau_{\nu}(r_{\rm outer})$
by using
the  ``vertically-averaged''
equations which give the
radial disk structure (Shakura \& Sunyaev 1973).
In qualitative terms, the viscous time
scale depends on the dimensions of the disk,
on the amount of stored gas, and on the viscosity
parameter $\alpha$.
Eqn. [A6] of Cannizzo \& Reiff (1992)
gives the midplane disk temperature $T=T(\Omega, \Sigma, \alpha_{\rm hot})$,
where the Keplerian frequency $\Omega(r) = (GM_{\rm WD} r^{-3})^{1/2}$
and $\alpha_{\rm hot}$ is the value of alpha in the outbursting
disk ($\simeq 0.1$, Smak 1984, 1998).
If we adopt an opacity law for the  
     ionized state $\kappa = 2.8\times 10^{24}$
g cm$^{-2}$ $\rho T^{-3.5}$
and use the condition of local hydrostatic equilibrium
$\Omega^2 h^2 = c_S^2 =  R T \mu^{-1} $ (where
$c_S$ is the local sound speed,  $h$ is the local
disk semi-thickness, $R$ the ideal  gas constant, and $\mu$
the mean molecular weight  $=0.67$),
we then get for the viscous time

$\tau_{\nu} \equiv (\alpha\Omega)^{-1} (r/h)^2 = 26$ d 
$m_1^{5/14}$  $(r_{10}/4.5)^{13/14}$  $(\alpha_{\rm hot}/0.1)^{-8/7}$
$(\Sigma/6\times 10^2$ g cm$^{-2})^{-3/7}$,

\noindent where $m_1=M_{\rm WD}/\msun$
and $r_{10} = r_{\rm outer}/(10^{10}$ cm).
The viscous time scale $\tau_{\nu} $
represents an $e-$folding time over which
    local surface density
 perturbations at a given radius
are smoothed out in a dynamically
evolving disk.
 This scaling predicts that if the depletion
of disk mass were large enough,
one should see a flattening
in the decay light curve, given that
 $\tau_{\nu} \propto \Sigma^{-3/7}$.

\section { Data and Modeling } 
\subsection { Data }

We examined the long term
AAVSO light curve of U Gem from 1908
to 2002 which consists of over 51,000
observations in the AAVSO
 International Database
from observers worldwide.
Figure 1 presents the frequency histogram
distribution for the outburst durations,
 measured from $m_V=12$
on the rising and declining portions of outburst.
 U Gem lies in the plane of the ecliptic
so that there are yearly sun gaps in the data.
   We omit from our analysis outbursts
  that straddle a sun gap, or that contain
any gap of more than 4 d.
 As with other DNe, one 
can see the characteristic bimodal distribution of
outburst durations. 
 Figure 2 shows the $m_V$ data from the AAVSO
of U Gem during a 140 d span beginning at JD 2446320
(i.e., 12.5 Sep 1985).
This interval covers the $\sim35$ d of the outburst,
as well as the following short outburst.
 The duration of the Oct 1985 outburst exceeded
by about a factor of 2 the second longest one.
 The two straight lines have slopes of 26 d mag$^{-1}$
and 1.4 d mag$^{-1}$, respectively.
        According to theory these
should represent the viscous and thermal
time scales, which differ by $\sim h/r$. 
  The combination of the finite scatter
in the data points and the limited dynamic
range of variation in $m_V$ on the slow part of 
the decay preclude one from being able to assess
a possible deviation from exponentiality 
in the slow decay light curve
as the disk drains onto the WD.
 This effect can be seen in the SU UMa superoutbursts
(e.g., Cannizzo 2001b for WZ Sge).

\subsection { Time Dependent Modeling}

As a final check, we utilize
our time dependent accretion disk code
to calculate a detailed light curve
(for descriptions of the code see
Cannizzo 1993b, 2001a).
   One recent upgrade to the code has been
the provision for a variable outer disk radius,
which has been implemented as described in
    Buat-Menard et al. (2001).
    One sees a rapid expansion of the
outer disk in outburst due to the outward
movement of hot, ionized gas in the heating
front, followed by a slow contraction in
quiescence due to the addition of low 
angular momentum material at the outer edge.
In addition, the disk mass in outburst is
generally larger, which leads to a slightly
brighter outburst.
  We assume $M_{\rm WD} = 1.1\msun$,
$r_{\rm inner} = 5\times 10^8$ cm, and
$r_{\rm outer} = 4.5\times 10^{10}$ cm.
To calculate $m_V$ we assume $D=96$ pc.
We perform a long run taking a mass transfer
rate ${\dot M}_T = 6.3\times 10^{18}$ g s$^{-1}$,
  and wait until the initial transients
associated with the initial surface density profile
     have disappeared.
  We then extract a long outburst from this run.
 Figure 3 shows the light curve, disk mass, and
accretion rate onto the WD for this model.
   We have not performed any special tuning
to enhance the duration of this long outburst,
therefore its duration is somewhat shorter
than the (unusually long) outburst of current interest ($\sim20$
d versus $\sim35$ d).
The two parallel dotted lines in the first panel indicate
a decay time of 26 d mag$^{-1}$. This slope matches well with
that of the computed light curve.
Also, the variation in disk mass indicated in the second
panel is in accord with the previous estimate of $\sim10^{24}$ g
for the accreted mass.

\section {Discussion and Conclusion }

Our result supports the standard model
  for dwarf nova outbursts in which
the outbursting disk decays primarily
via simple accretion onto the WD when 
in the ``viscous plateau'' (Cannizzo 1993b) 
 stage, and $\alpha_{\rm hot}\simeq 0.1$.
Of all outbursts in U Gem 
   observed by the AAVSO 
over the past century, only  the Oct 1985 outburst
had a long enough duration to allow a 
fairly precise determination of its viscous
plateau decay time scale.
  The decay time $\sim26\pm6$ d mag$^{-1}$
is about $\sim2-5$ times longer than observed
in, for instance, the 2001 superoutburst 
           of WZ Sge (Cannizzo 2001b),
and is consistent with what one would predict based
on the orbital period and outburst energetics.
  The more normal state of affairs for long period DNe 
in which 
  one frequently sees flat-topped outbursts
is that 
   the short duration ($\sim7-10$ d) 
 of the viscous plateau, 
  combined
  with the 
  slow viscous time in the outer
disk,
conspire to produce a much smaller
dynamic variation in $m_V$.
In fact the variations that are 
usually seen may be
due almost entirely 
      to sloshing action of the gas in
the disk as it responds
to the matter redistribution accompanying the outburst,
therefore 
one cannot reliably determine a slope for the
viscous plateau.
DNe 
 appear to be unique in that the identification
of viscous and thermal decays is reasonably firm.
Physical effects extrinsic to the underlying accretion
disk limit cycle theory such as strong irradiation
and evaporation do not alter the basic picture in
a fundamental way.   
  In contrast, the identification of viscous decays
with outbursts seen in the X-ray novae - interacting
binaries in which the accretor is a black hole or neutron
star - is much more problematic 
(Ertan \& Alpar 1998,
 King \& Ritter 1998, 
 Shahbaz, Charles, \& King 1998,
   Cannizzo 2000,
 Wood et al. 2001).

We sincerely thank variable star
observers worldwide who, for more
than 90 y, have
contributed observations
on U Gem to the AAVSO International
Database.
It is their observations that make this study possible.

\vfil\eject
\centerline{ FIGURE CAPTIONS }
%\medskip

Figure 1. Frequency
 histogram distribution of all outburst
durations in U Gem, compiled from the long term
AAVSO light curve. 
 We utilize 1 d bins, and measure the times
from $m_V=12$ on the rising and declining portions
of the outbursts.
The total number of outbursts
in the distribution is 78,
with 31 below $t_{\rm outburst}=9$ d,
and 47 above.
As with SS Cyg,
the distribution is strongly bimodal, with 
peaks at $\sim5$ d and $\sim12$ d.
 The 39 d duration of the Oct  1985 outburst is
indicated by the arrow.

Figure 2. 
 AAVSO data of U Gem spanning 140 d, starting from
JD 2446320.
 The two straight lines have slopes of 26 d mag$^{-1}$
and 1.4 d mag$^{-1}$, respectively.

Figure 3. Time variation during a computed
``long'' outburst of the apparent visual
magnitude ({\it top panel}), the disk mass ({\it middle  panel}),
and the rate of accretion onto the WD ({\it bottom panel}).
The two dotted lines in the first panel indicate a rate
of decay of 26 d mag$^{-1}$.

%2445619 12.97996
%2445718 5.977156
%2446383 39.15527
%2446438 3.850909
%2446568 4.193646

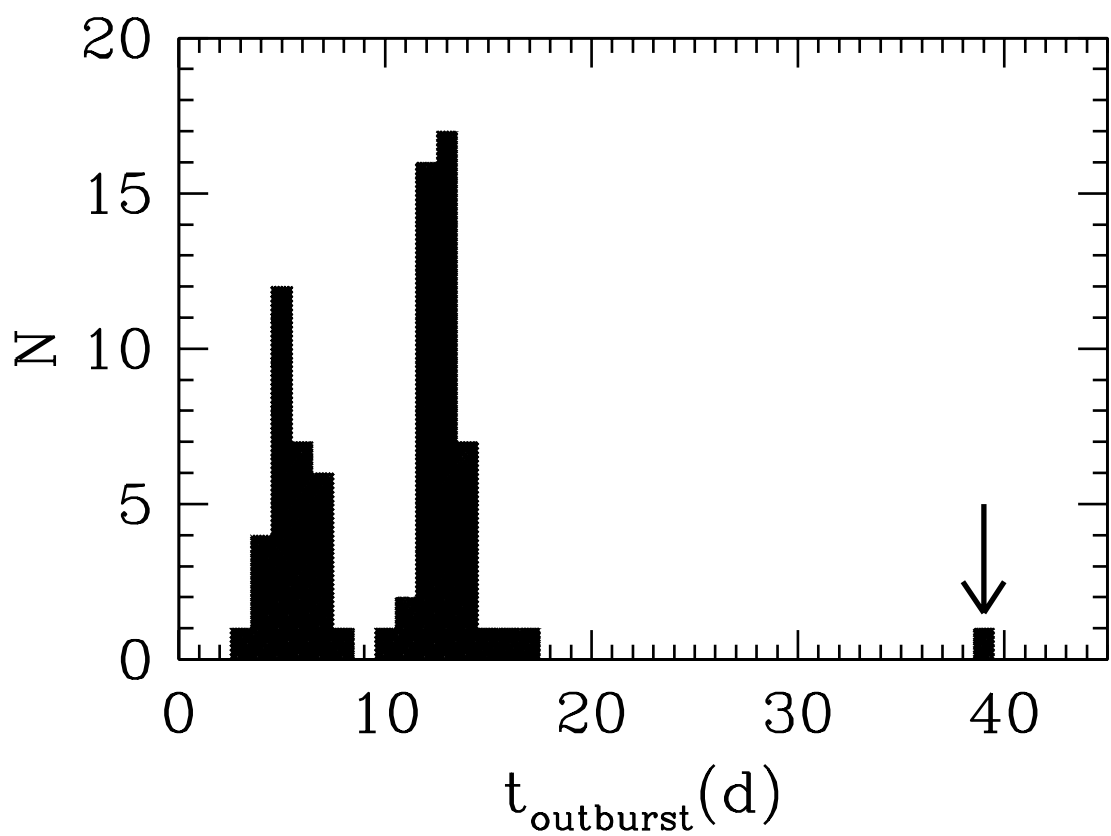
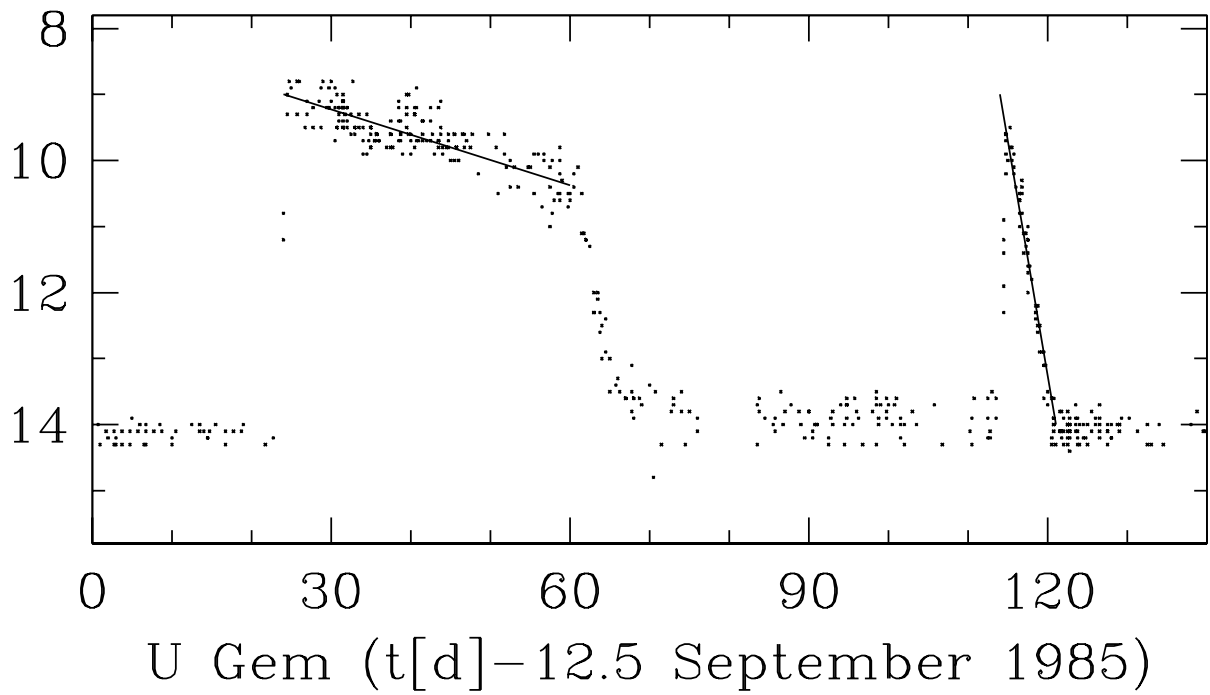
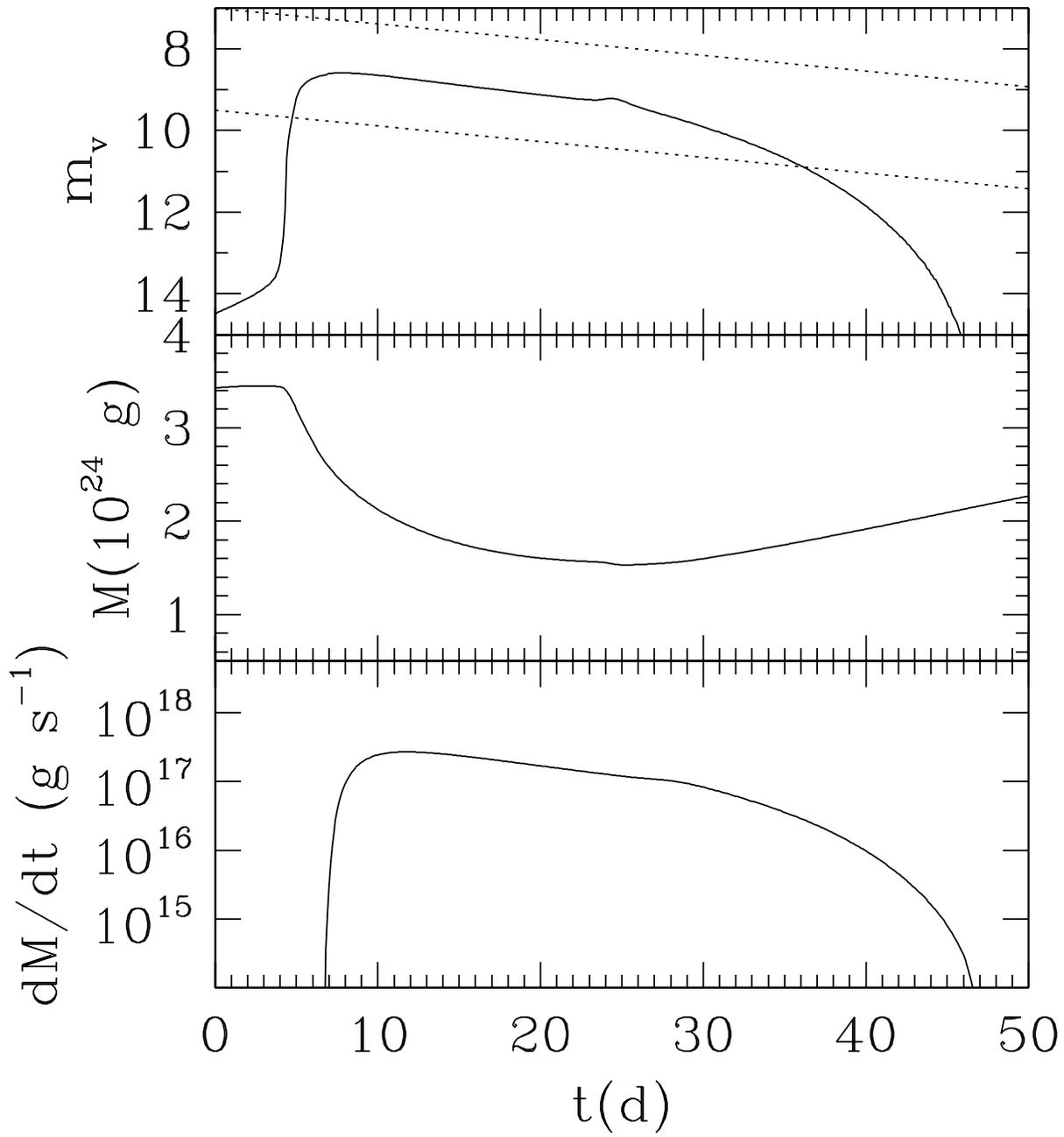
\end{document}